%% file: papeth.tex
\documentclass[11pt]{article}
\usepackage{a4,moriond,epsfig,amssymb,here}

\def\PLB{{\em Phys. Lett.}  B}

\def\be{\begin{equation}}
\def\ee{\end{equation}}
\def\bea{\begin{eqnarray}}
\def\eea{\end{eqnarray}}

\include{mydefs}
\begin{document}
\begin{flushright}
{\large\bf ETHZ-IPP PR-99-03\\ 
May 31st, 1999}
\end{flushright}
\vspace*{5cm}
\title{\Large\bf  The search for Higgs particles at LEP}
\author{\large\bf M. Felcini}

\address{\large Swiss Federal Institute of Technology, \\
Institute for Particle Physics,\\
Z\"urich, Switzerland}

\maketitle
\abstracts{\large
The results of the experimental searches for Higgs particles at LEP, 
using the data collected at centre-of-mass energies up to 189 GeV, 
are reviewed and the prospects for the near future outlined.
}
\vspace*{5cm}
\begin{center}
Talk given at
the XXXIVnd Rencontres de Moriond,\\ 
ELECTROWEAK INTERACTIONS AND UNIFIED THEORIES, March 13--20 1999,\\
to be published in the Proceedings. 
\end{center} 

\newpage
\section{Introduction}
The Standard Model (SM) of the electroweak interactions has been subject to 
very precise tests at LEP and elsewhere. Up to now its prediction have been 
found in very good agreement with the experimental observations. 
\eg\ the masses and couplings of the vector bosons, W and Z.
To explain the origin of the W and Z boson masses, 
as well as  of the fermion masses, the model requires
the existence of a weak isospin doublet of 
complex scalar fields which is responsible for the spontaneous breakdown of 
the gauge symmetry. Spontaneous symmetry breaking is the mechanism which 
generates masses for the bosons and the fermions. A direct consequence of 
this mechanism is the existence of one physical neutral scalar, the Higgs 
boson. 
The mass of the Higgs boson is not predicted by the theory. There
are theoretical indications that the existence of a relatively light 
Higgs would indicate the presence of new physics in the few TeV 
range~\cite{lep2yebo}.

It is generally accepted that, despite the great success of its electroweak 
predictions, the SM may be an effective theory valid only at low energies. 
Indeed, several fundamental questions are raised 
which cannot get a satisfactory 
answer in the context of the SM. Supersymmetry offers solutions to some of 
these problems. In its simplest realisation, the minimal supersymmetric 
extension of the SM (MSSM), there are two Higgs doublets resulting into 
five physical Higgs particles:
two neutral CP-even, h and H ($\Mh<\MH$), 
one neutral CP-odd, A, and two scalar charged particles, $\Hp$ and $\Hm$. 
Their masses at 
tree-level depend only on two MSSM parameters, \eg\
the ratio of the vacuum expectation value  of the two Higgs fields, $\tanbet$, 
and the mass of the CP-odd Higgs boson, $\MA$.   
When introducing radiative corrections, 
the Higgs masses and couplings (which determine cross sections and
decay branching ratios) become dependent 
also on the top mass, the scalar top mass and other parameters 
of the model. 
In particular, the dominant radiative corrections to the Higgs 
masses grow with the fourth power of the top mass and they are logarithmically
dependent on the stop mass. 
The mass of the CP-odd Higgs boson can be as high as 
$\sim 1$ TeV, the mass of the heavier CP-even Higgs boson should be typically 
higher than $\sim 120$ GeV and the mass of the charged Higgs typically 
higher than $\sim$ 90 GeV~\cite{lep2yebo}. 
Differently from the SM, in the MSSM an upper limit on the mass of the 
lightest CP-even Higgs 
boson mass $\Mh$ can be set, 
which, depending on $\MA$, $\tanbet$ and the mixing in the stop sector,
can be as high as $\sim$130 GeV~\cite{lep2yebo}, for a top mass 
value in the measured range $\Mt=174.3\pm 5.1$ GeV~\cite{heinson99}. 
There are  
several scenarios in which this Higgs mass upper bound is $~100$ GeV or less, 
of particular interest for the LEP searches.  

Non supersymmetric extensions of the SM are also advocated to solve some of 
the problems left open by the SM. They predict a different Higgs 
phenomenology from the one predicted by the SM and the MSSM. These predictions
can also be tested experimentally.    


There are two main sources of experimental informations about 
Higgs bosons: on one hand, the precise measurement of the electroweak 
parameters and, on the other hand, the direct searches.
In the SM (and the MSSM) the electroweak observables are sensitive, 
through loop corrections, to the 
Higgs mass, $\MH$, though only logarithmically. 
The increasing precision of present 
experimental data from LEP, Tevatron and SLC, makes it possible to 
derive significant upper bounds on the SM Higgs 
(and the lightest neutral MSSM Higgs) 
mass, which are of great interest, and encouragement, for the direct searches.
A global fit to all the available measurements of 
electroweak observables gives  an upper bound~\cite{lemo99ew}
$\MH<220$ GeV at 95\% CL.
The central value with the 68\% CL range indicated by the 
fit~\cite{lemo99ew} is:
$\MH=71^{+75}_{-42}+5$ GeV (the limit from direct searches, see below, is not 
included in the calculation). 

A wide variety of experimental informations  
comes from the direct Higgs searches, devoted to detect direct
evidences of Higgs particles, independently of model assumptions.
With this aim, comprehensive searches are being made at LEP, covering 
a large number of decay channels and signatures. 
The discovery of (at least one of) the Higgs boson(s) is, indeed, one of the 
main goals of the LEP programme.  
The centre-of-mass energy ($\sqrts$) and corresponding 
luminosity of the LEP collider 
have been constantly increased since the end of 1995. 
In 1996 about 20 \invpb\ where collected by each LEP experiment at
$\sqrts=$161--172 GeV. In 1997 the $\sqrts$ was raised to 183 GeV 
and about  60 $\invpb$ were collected per experiment. 
The direct searches for Higgs bosons at LEP at  
$\sqrt{s}\leq$ 183 GeV have resulted in several lower
limits on the masses of Higgs particles.
The combined LEP limits~\cite{lemo99pp} are: 
for the SM Higgs, $\MH> 90$ GeV (individual limits in the range 
$\sim$86--88 GeV), 
for the MSSM neutral Higgs masses $\Mh$ and $\MA> 80$ GeV 
(individual limits in the range $\sim$68--76 GeV), for the charged 
Higgs $\MHpm > 70$ GeV (individual limits in the 
range $\sim$57--59 GeV). 
      
The LEP data collected from May to November 1998 at $\sqrt{s}=189$ GeV
and amounting to approximately $170\ \invpb$ per experiment 
give the possibility 
to explore higher Higgs mass ranges.     
LEP is at present the most powerful tool for the direct search of 
Higgs particles in mass regions never explored before.

\section{The Higgs boson of the Standard Model}
Within the framework of the SM, the Higgs is dominantly produced via the
Higgs-strahlung process~\cite{lep2yebo}. 
Additional production processes like W or Z fusion 
contribute very little (less than 1\%) to the Higgs production rate in the 
mass range of interest for the experimental search. 
The SM Higgs production cross section at $\sqrt{s}=189$ GeV is 0.34 pb and 
0.178 pb for Higgs mass of 90 and 95 GeV, respectively. 
For comparison, 
the predicted cross sections of  
other SM processes, producing the main background to the Higgs 
search, are: $98$ pb for $\EE\to\QQ$ production, 
$16$ pb for WW production and 
0.62 pb for the ZZ production. 
In particular, the ZZ process in its
$\rm\BB X$ final states provides the ``irreducible'' background to the Higgs, 
in the Higgs mass region around 90--95 GeV.
The measured cross section values 
for these processes are in very 
good agreement with the SM predictions \cite{cavallari99}.   

In the mass range of interest at LEP, the SM Higgs decays dominantly into 
into $\BB$ and $\TT$ pairs 
with branching ratios $\simeq 83\%$ and $\simeq 8\%$, respectively. 
There are four final states of interest for the HZ search:
final state with four jets, $\BB\QQ$, and 60\% relative rate, 
with jets plus missing energy, 
$\BB\NN$, and 18\% relative rate, 
with jets plus electron or muon pairs, 
$\BB\EE$ or $\BB\MM$, with 
6.6 \% relative rate,  and with 
jets plus tau pairs, $\BB\TT$ and $\TT\QQ$, with 3.3\% and 5.6\% 
relative rate, respectively.
Thus,    
the identification of b-quarks, commonly referred to as b-tagging, is an 
indispensable tool for an efficient Higgs search, as very powerful  
mean to reduce backgrounds containing light (non b) quarks.

The four LEP experiments use to combine the b-tagging information with 
event shape and kinematic variables into a global discriminant quantity
(\eg\ likelihood, neural network output or other) \cite{nuhgmo99}.
This quantity is a measurement of the ``Higgs-likeness'' 
of an event: background events are preferably distributed  at  low values 
while Higgs events have preferably large value of ``Higgs-likeness''. 

The total number of candidates (events with large value of the global 
discriminant)  
selected by the four experiments are reported in Tab.~1 
and compared to the expectation from SM background processes 
and to the expected contribution of a 95 GeV SM Higgs signal. 
No significant excess of events is observed by any of the 
experiments over the expected background.
The reconstructed Higgs mass distributions of the most significant
candidates selected by the four LEP experiments are shown 
in Fig.~1. A slight excess of events over the expected background 
is observed by OPAL in the mass region around the Z mass. 
The LEP combined mass distribution \cite{mass} is shown in Fig.~2,
where the combined data (full dots) are compared to the total 
expected background (solid line) and to the sum of the background
plus the expected signal for a 95 GeV SM Higgs (dashed line). 
Good agreement is observed between data and SM expected 
background. The largest ``discrepancy'' is observed in the mass bin 
92-96 GeV with 47 events observed and 37.5 expected from SM background
(see last line of Tab.~1). For comparison,  a 95 GeV SM Higgs would 
contribute 24.6 events in this mass bin.  

From the distributions (\eg\ in mass) of the observed candidates 
and of the expected  signal and background, it is possible to evaluate, 
for a given Higgs mass $\MH$,  
the probability $CL_s$ that the Higgs signal be
undetected in the observed data. 
Then, a Higgs signal of mass $\MH$ is excluded with  
a confidence level $CL=1-CL_s$.
The $CL_s$ curves as function of $\MH$  
are shown in Fig.~3 for the four experiments \cite{nuhgmo99}. 
The 95\%CL lower limit on the Higgs mass is 
the value of $\MH$  for which the $CL_s$ is equal to 0.05.

Together with the observed $CL_s$, calculated using 
the {\bf observed} distribution,
the four experiments report  
the expected $CL_s$ \cite{nuhgmo99}, calculated using the {\bf expected} 
background distribution.  
The expected $CL_s$ 
is the average $CL_s$ which would be measured if the 
experiment could be repeated a very large number of times, 
in the hypothesis that no signal, but only background,  would contribute
to the observed candidates. 
The expected $CL_s$ is used by the experiments 
to optimise the sensitivity of the analysis: 
more performant analysis should result, for a given $\MH$, into a lower $CL_s$,
 or, in turn, for a given $CL_s$, into a higher $\MH$ limit.   

The observed and expected limits on $\MH$ 
and the probability to obtain a lower
Higgs mass limit than the one actually observed  are summarised in Tab.~2 for 
the the four experiments. 

\section{Higgs bosons in the Minimal Supersymmetric Standard Model}
The production of h and A at present LEP energies is expected to occur 
through two processes: the  Higgs-strahlung, $\EE\to\hZ$, 
and the associate hA production, $\EE\to\hA$. 
The cross sections for these processes are given by \cite{lep2yebo}
$\sigmahZ=\stwoab\sigmaHZSM$ and $\sigmahA=\ctwoab\bar{\lambda}\sigmaHZSM$, 
where $\sigmaHZSM$ is the SM Higgs production cross section, 
$\bar\lambda$ is a kinematic factor depending on $\Mh$, $\MA$ and $\sqrts$, 
while $\alpha$ is the mixing  angle of the two CP-even Higgs fields.
Due to the factors $\stwoab$ and $\ctwoab$, the two Higgs production 
processes, hZ and hA, are complementary.
At low $\tanbet$,  where $\stwoab$ is large, 
hZ production is dominant and 
practically the MSSM Higgs phenomenology reduces to the SM one, with similar 
cross sections and decays, except if $\Mh>2\MA$. Then  $\h\to \A\A$ decay
is dominant and the SM Higgs signatures $\BB$Z are replaced by $\BB\BB$Z. 
At large $\tanbet$,  where $\ctwoab$ is large,  
hA production is dominant. As an example the hA cross section 
for $\Mh=\MA=80$ GeV and $\tanbet=50$ is 75 fb while the hZ cross section 
is approximately 6 fb. 
In the intermediate region, for $\tanbet$ between approximately 2 and 5, 
the two processes give both sizeable contribution to the observable signal.

The dominant decays are expected to be $\rm h,A\rightarrow b\bar b$ 
and $\h\A\to\TT$ with branching ratios approximately 90\% and 7\%, 
respectively.
Signal topologies considered 
for hA search are 4 or 6 b-jets 
or 2 isolated taus plus 2 b-jets.    
The experimental search for hA is very strongly based on b-tagging
(even more than the SM-like hZ search),
though other quantities, such as topological and kinematic variables 
are also used in combination with the b-tagging informations,  to 
improve signal discrimination from background \cite{nuhgmo99}.

As no significant excess is observed in any of the channels investigated,
the results of the different searches are combined, using the same statistical
methods as for the SM Higgs search and interpreted in terms of 
excluded regions in the MSSM parameter space. 
Differently from the SM, where the expected number of signal events 
depends only on $\MH$ (at a given $\sqrts$), in the MSSM,
the expected number of 
signal events is calculated using cross sections and decay branching ratios 
which depend on the two masses $\Mh$ and $\MA$ and other MSSM parameters. 
The signal efficiencies also depend, in general,  
on both Higgs masses $\Mh$ and $\MA$. 
A given point of the 
MSSM parameter space is then excluded if the observed $CL$ is at least 95\%
for that parameter choice. 

The interpretation of the results of the MSSM searches is, 
generally, done in
the framework of the constrained MSSM, with a  
number of free parameters smaller than the general MSSM. 
The constrained model assumes a 
unified scalar fermion mass, $\MSUSY$, 
a unified gaugino mass, $\Mtwo$, and a unified   
sfermion tri-linear coupling, $A$ .  
The remaining parameters of the model are: the top mass $\Mt$, 
the SUSY Higgs mass parameter, $\mu$, 
$\tanbet$ and $\MA$.
The experiments use to present their results as excluded regions in the 
planes [$\Mh$,$\tanbet$],  [$\MA$,$\tanbet$] or [$\Mh$,$\MA$].
The other parameters of the model are either  fixed 
(by ALEPH, DELPHI and L3)  to the two set of 
values determining  maximal mixing or no mixing in the stop 
sector \cite{lep2yebo}, 
or a scan is performed (by OPAL) over these parameters,
aiming to evaluate the effect of their
variations on the excluded Higgs mass and $\tanbet$ regions \cite{nuhgmo99}.
    
Regions excluded using the data at 189 GeV, 
combined with the results from the previous runs at lower energies, are shown 
in Fig. 4: for L3, in the [$\Mh$,$\tanbet$] plane and  for DELPHI,
in the [$\MA$,$\tanbet$].
The lower limits on the Higgs masses  and the excluded $\tanbet$ ranges for 
maximal and minimal mixing assumptions are summarised in Tab.~3 for the four 
experiments.

\section{Higgs particles beyond Minimal SUSY}
\subsection{Charged Higgs}
Charged Higgs bosons are predicted by all models beyond the SM 
with more than one Higgs doublet. 
As mentioned in Introduction, the charged Higgs in 
the MSSM is expected to be
heavier than 90 GeV in most of the 
MSSM parameter space when radiative corrections are included.
Thus, if LEP found a  charged Higgs with mass below the W mass this would 
very strongly constrain the MSSM parameter space.  

Charged Higgs bosons are produced in pair via the s-channel process
$\EE\to\HpHm$. The cross section depends only on the charged Higgs mass,
$\MHpm$, 
and $\sqrts$. As an example, at $\sqrts=189$, the 
cross section for $\MHpm=70$ GeV is expected to be 0.25 pb. 
In the mass range accessible to LEP searches, the charged Higgs is expected 
to decay into a pair of quarks or into $\tau\nu$. The decay branching 
ratios are model dependent.

The search is done, independently of the decay branching ratios,  
in all three possible $\Hp\Hm$ final states:
$\tau\nu\tau\nu,\ 
\tau\nu q\bar{q}'$ and $\rm q \bar{q}' q''\bar{q}'''$.   
The main background in all three search channels comes from WW production. 
In the four-jet channel  
additional background comes from $\rm q\bar q$ events 
with hard gluon emission. 
The details of the analyses can be found in 
Ref. \cite{chhgmo99}. 

As no clear signal has been observed, limits on the charged Higgs mass 
as function of the branching ratio in $\tau\nu$, $BR(\Hpm\to\tau^\pm\nu)$, 
are derived. 
They are reported in Fig.~5 for ALEPH and OPAL. 
In Tab.~4, the lower limits on $\MHpm$ set by the four experiments, 
are reported for 
$BR(\Hpm\to\tau^\pm\nu)$ equal to 1, 0 and independent of the decay branching 
ratio. This last one is the most conservative lower limit on $\MHpm$, 
valid for any value of $BR(\Hpm\to\tau^\pm\nu)$.  

\subsection{Invisible Higgs decays}
The LEP experiments have also searched for hZ production with 
the Z decaying into quarks and leptons and the Higgs decaying into 
invisible particles. Invisible Higgs decays are possible in the framework 
of supersymmetric and non supersymmetric models.
In the former case, the Higgs decays into a pair of lightest neutralinos,
assumed to be stable and undetectable.
In the latter case, the invisible particles can be, for example, 
Majorons \cite{inhgmo99}. 

The experimental search is based on topological and kinematic characteristics 
of the signal events. After a number of cuts on the most discriminating 
variables, or on a global discriminant, combination of them, 
the distribution of the reconstructed missing mass, recoiling against the
visible system, after constraining the visible mass to the Z mass, is
inspected for any excess over the expected background.
 
As no significant excess is observed, a CL calculation is used to derive 
upper limits on the rate of invisible Higgs decays as function of the Higgs
mass. 
Assuming hZ production cross section equal to the SM one 
and 100\% branching ratio into 
invisible particles, the observed lower limit  on the Higgs 
mass is 92.8 GeV from ALEPH \cite{inhgmo99}.
The Higgs branching ratio into invisible particles, 
BR$BR\rm(\h\to inv)$, is actually a 
free-parameter, which 
can assume any value between 0 and 1, depending on the model considered. 
The hZ production cross section, $\sigmahZ$,  is also model-dependent. 
Thus in a model-independent approach, limits on the neutral Higgs mass
can be derived as function of $BR\rm(\h\to inv)$ and $\sigmahZ$. 
The results of the searches for visible (SM-like)
and invisible Higgs decays can be combined, and the excluded region in the
[$\Mh$,$BR\rm(\h\to inv)$] plane can be determined for any value of 
$\sigmahZ$. 
Assuming for $\sigmahZ$ the SM value, a lower limit on $\Mh$ of 90.5
GeV is found, independent of $BR\rm(\h\to inv)$, as shown 
in Fig.6(left) from DELPHI \cite{inhgmo99}.

\subsection{$H\gamma$ production, $H\rightarrow\gamma\gamma$ decays }
Search for Higgs bosons have also been done in events with one, two
or three photons \cite{opgg98pa,gghgmo99,degg99ep}.
These final states can occur via the processes: $\EE\to\Hv\gamma\to\BB\gamma$, 
$\rm\EE\to\Hv\Z\to\gamma\gamma X$ and 
$\EE\to\Hv\gamma\to\gamma\gamma\gamma$. In the framework of the SM,
these processes have very small rates, making them practically 
undetectable at LEP.
However, in models beyond the SM, e.g with anomalous couplings of the 
Higgs boson to photons and Z \cite{gghgmo99,degg99ep} or with fermiophobic
Higgs bosons \cite{opgg98pa,gghgmo99,degg99ep}, 
the $\EE\to\Hv\gamma$ cross section and the \
$\Hv\to\gamma\gamma$ decay rate can be largely enhanced 
(few order of magnitudes) with respect to the SM expectations, such that 
their signals could actually be observed at LEP.

No signal has been observed in any of the channels investigated, thus
upper limits have been set on the rates of the different processes between 
few 10 and few 100 fb for Higgs masses up to 180 GeV.
An example is given in Fig.~6(right) from L3 
showing the experimental upper limit on 
the rate for $\HZ\to\gamma\gamma\gamma$,
$\rm\sigma(\Hv\gamma)\times BR(\Hv\to\gamma\gamma)$, compared to the  
expected rates in the presence of anomalous Higgs 
couplings at $\sqrts=$ 189 GeV \cite{gghgmo99}. 
The expected rates are plotted for 
different values of $\Lambda$, the typical energy scale at which  
new interactions should take place. 
\section{Conclusions}
The data collected at $\sqrts=$ 189 GeV, corresponding to a luminosity 
of approximately 170 \invpb\  
per experiment, have opened new opportunities to search for Higgs bosons
at LEP. 
The search for 
the SM Higgs boson has been performed in the mass range up to 95 GeV.   
Pair produced MSSM neutral Higgs boson have been searched for 
in the mass range 
up to $\sim$80 GeV and charged Higgs bosons  up to 75 GeV. 
Non-standard Higgs decays and production mechanisms have also been 
investigated,  as decays into invisible particles or into 
photons and $\Hv\gamma$ production. 
No significant excess is observed (yet), compared to the expected 
contribution from known SM physics processes,  
in any of the channels investigated.


In the next few months the centre-of-mass energy of the LEP collider
 will be raised up to 
196--198 GeV. The luminosity expected to be delivered to each experiment 
is similar to the one of last year, between 150 and 200/pb. 
With such an amount of data we will be able to search for 
a Higgs boson with a mass up to 100 GeV.  
An additional similar amount of luminosity is expected to be delivered 
next year (2000) at $\sqrts=$200 GeV. 
This should allow to explore the mass range up to 
105 GeV.  
Rather than with a list of limits from the previous searches, I prefer to 
conclude with this look at the  (very near) future 
full of hope for the final realisation of our `Higgs dreams'!

\input{refere}
\input{figure}
\end{document}

%% file: mydefs.tex
\def\eg{{\it e.g.}}

\newcommand{\EE}{\rm e^+ e^-}
\newcommand{\MM}{\mu^+ \mu^-}
\newcommand{\TT}{\tau^+\tau^-}
\newcommand{\QQ}{\rm q \bar{q}}
\newcommand{\BB}{\rm b \bar{b}}

\newcommand{\NN}{\nu \bar{\nu}}

\newcommand{\Z}{\rm Z}
\newcommand{\Hv}{\rm H}
\newcommand{\h}{\rm h}
\newcommand{\A}{\rm A}
\newcommand{\Hp}{\rm H^+}
\newcommand{\Hm}{\rm H^-}
\newcommand{\HpHm}{\rm H^+H^-}
\newcommand{\Hpm}{\rm H^\pm}

\newcommand{\HZ}{\rm HZ}
\newcommand{\hZ}{\rm hZ}
\newcommand{\hA}{\rm hA}

\newcommand{\sigmaHZSM}{\sigma_{\rm HZ}^{SM}}
\newcommand{\sigmahZ}{\sigma_{\rm hZ}}
\newcommand{\sigmahA}{\sigma_{\rm hA}}

\newcommand{\MH}{M_{\rm H}}
\newcommand{\Mh}{M_{\rm h}}
\newcommand{\Mt}{M_{\rm t}}
\newcommand{\MA}{M_{\rm A}}
\newcommand{\MHpm}{M_{\rm H\pm}}

\newcommand{\Mtwo}{M_{\rm 2}}
\newcommand{\MSUSY}{M_{\rm SUSY}}

\newcommand{\tanbet}{\mbox{$\tan{\beta}  $}}
\newcommand{\stwoab}{\mbox{$\sin^2{(\beta-\alpha)}  $}}
\newcommand{\ctwoab}{\mbox{$\cos^2{(\beta-\alpha)}  $}}
\newcommand{\invpb}{\mbox{$\rm pb^{-1}\ $}}
\newcommand{\sqrts}{\mbox{$\sqrt{s}$}}

%% file: figure.tex
\begin{table}[H]
\caption{ 
The number of SM Higgs candidates 
selected by the four LEP experiments,   
compared to the expected background and to the contribution from a 95
GeV Higgs signal, for any value of the reconstructed mass, 
and (last line) within the mass bin 92--96 GeV. 
} 
\bf
\begin{center}
\begin{tabular}{ l c c c  }  \\
\hline
                &Obs, & Exp. &  Exp. Signal\\
                &Data & Bkgd &  (\boldmath$\MH$=95 GeV)\\
\hline
   ALEPH  & 53 &44.8   &  13.8  \\
    DELPHI & 26 &31.3  &  10.1  \\
     L3     & 30 &30.3 &   9.9  \\
      OPAL  & 50 &43.9 &  12.6  \\
\hline  
Total    & 159 &  150 & 46.4\\
\hline
\boldmath$\rm\Delta \MH=$92--96 GeV   & 47 & 37.5 & 24.6\\
\hline
\end{tabular}
\end{center}
\end{table}

\begin{figure}[H]
\caption{
The distributions of the reconstructed mass for the 
SM Higgs candidates selected by 
the four LEP experiments \protect\cite{nuhgmo99}. 
} 
\begin{center}
\begin{tabular}{ l l  }  \\ 
\hspace*{-1.cm}
\epsfig{file=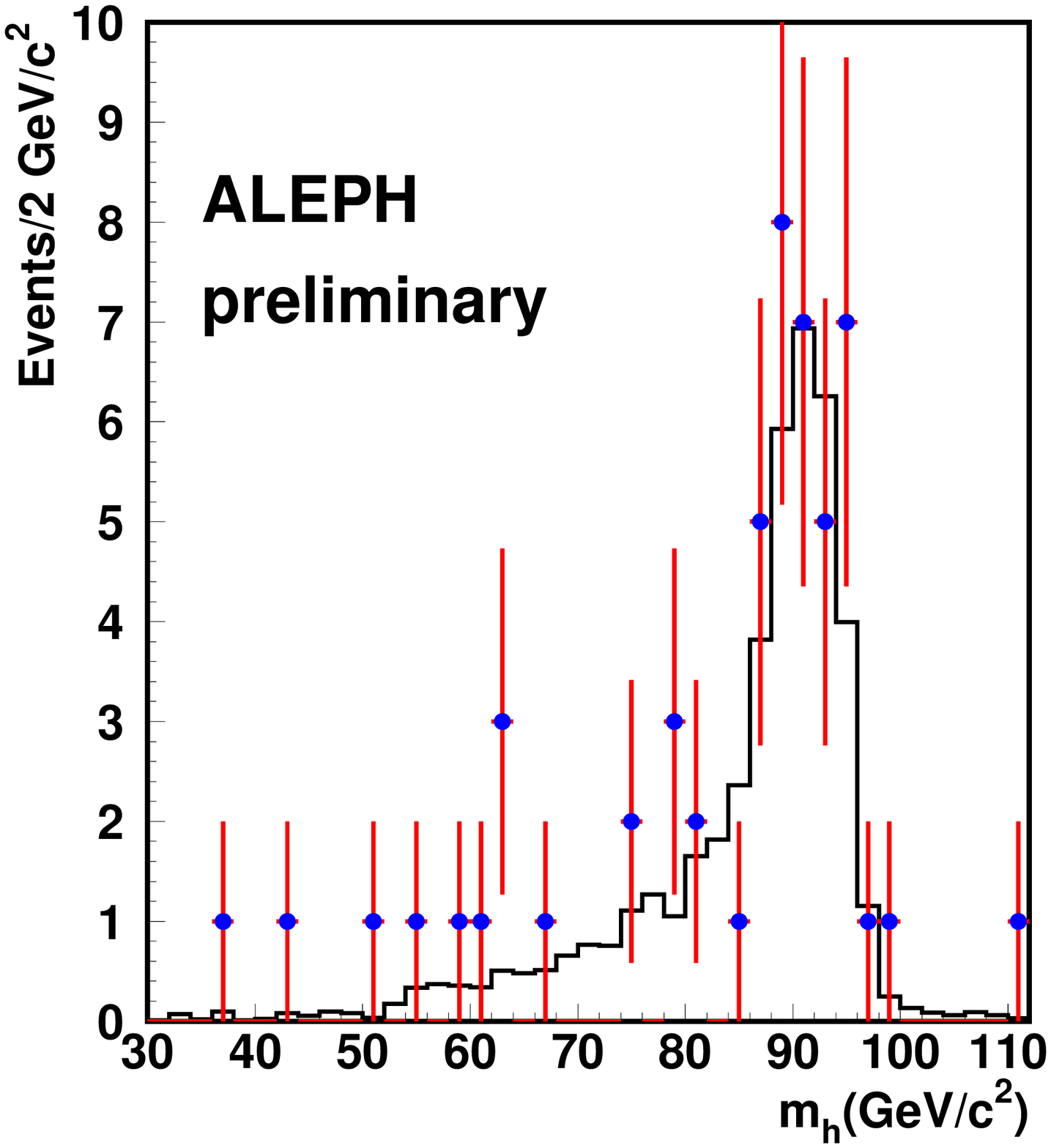,width=6.cm,height=6.5cm,clip=} &
\hspace*{-1.cm}
\epsfig{file=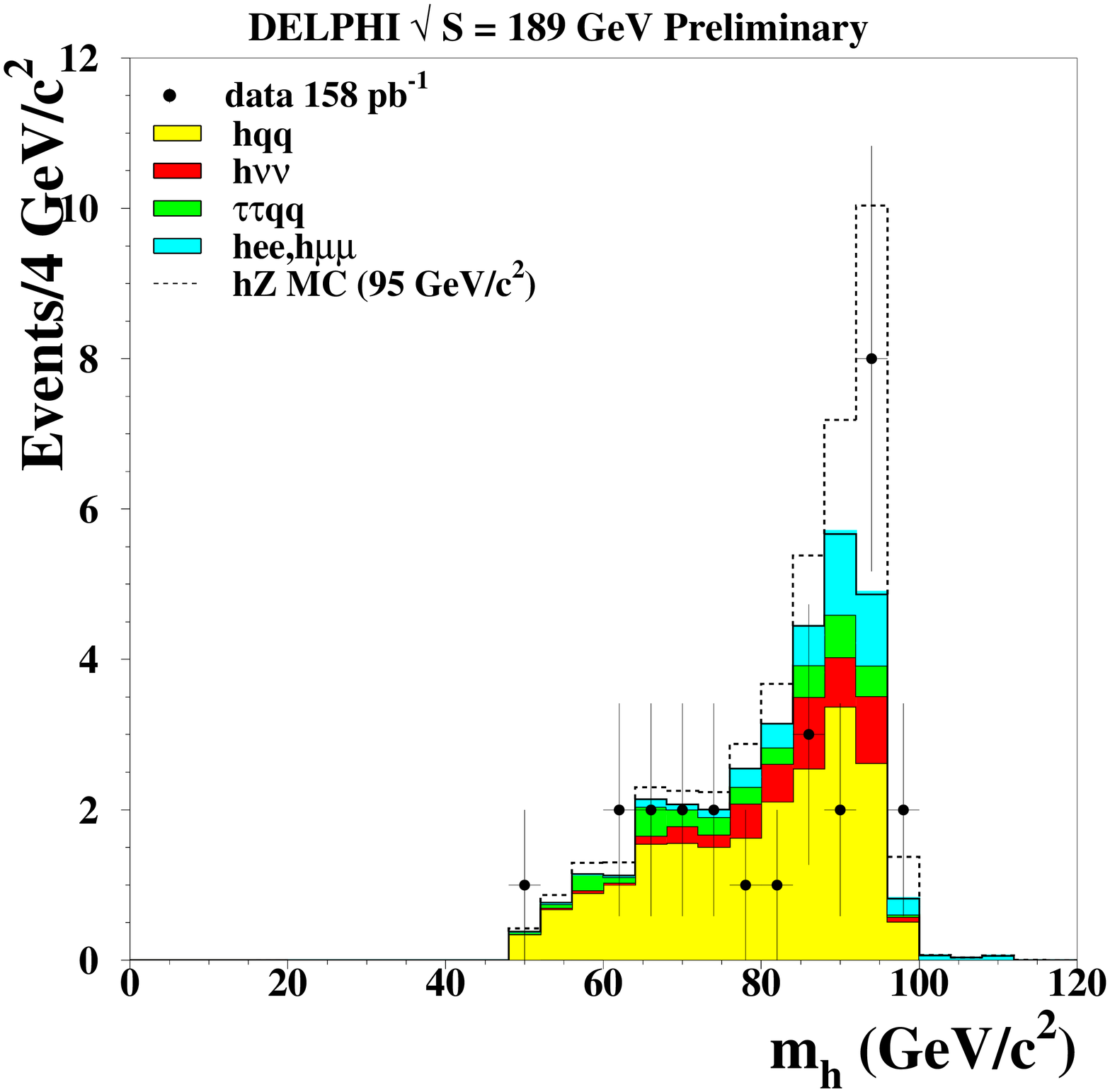,width=6.cm,height=6.5cm,clip=} \\
\hspace*{-1.cm}
\epsfig{file=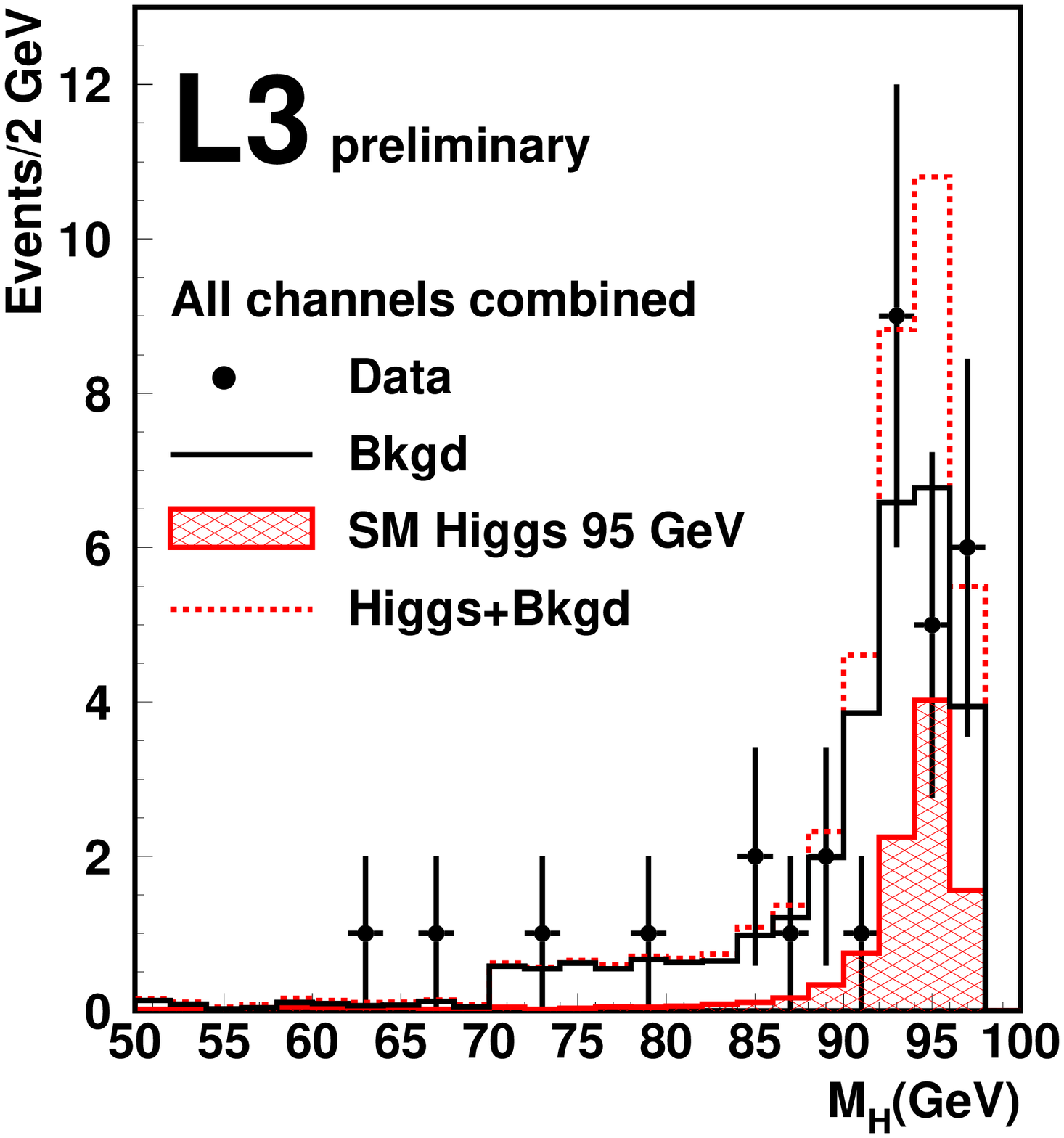,width=6.cm,height=6.5cm,clip=} &
\hspace*{-1.2cm}
\epsfig{file=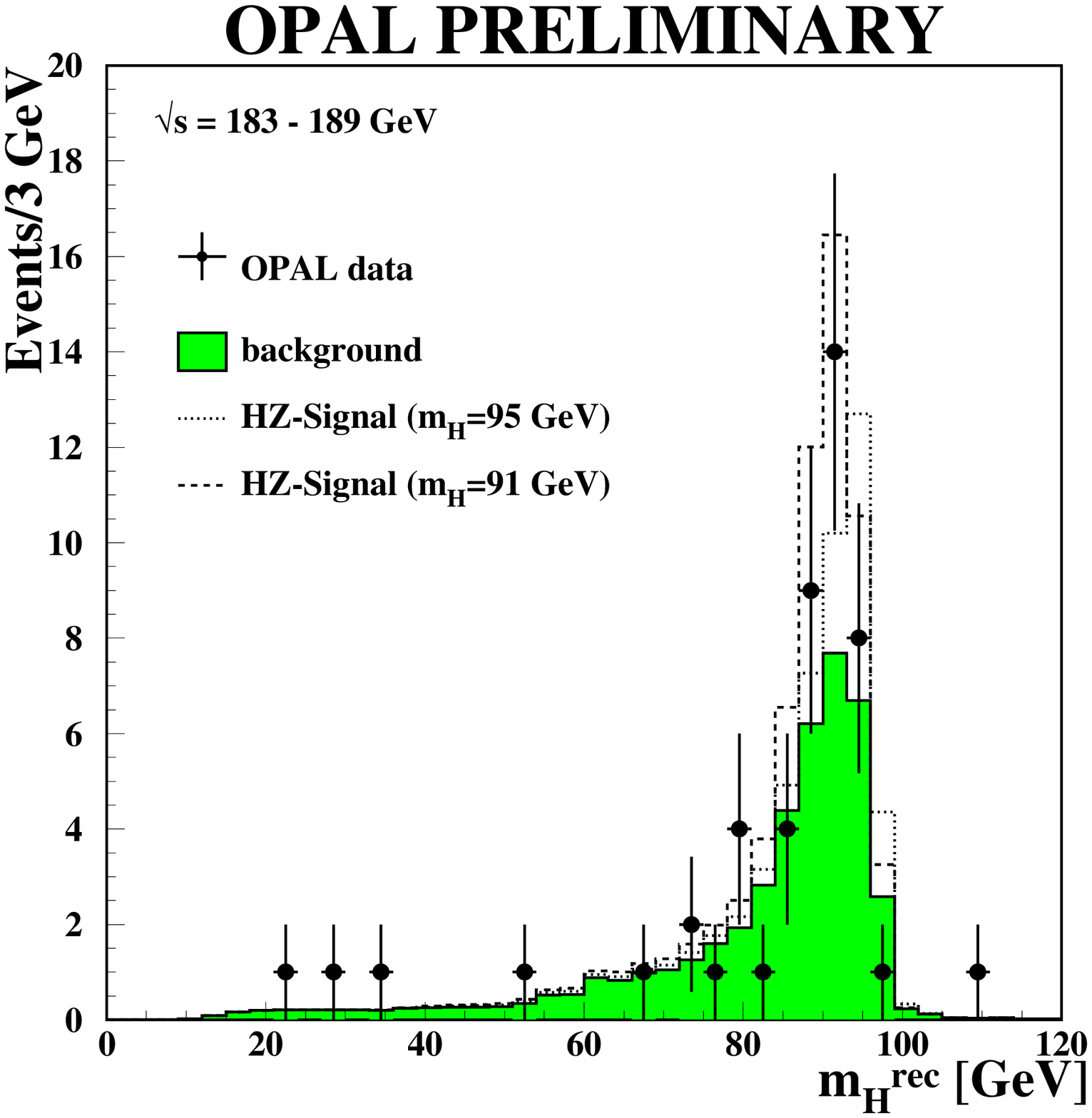,width=6.cm,height=7.5cm,clip=} \\
\end{tabular}
\end{center}
\end{figure}

\begin{figure}[H]
\caption{Combined LEP mass distribution \protect\cite{mass} 
of SM Higgs candidates selected by the four LEP experiments.
The combined data (full dots) are compared to the total 
expected background (solid line) and to the sum of the background
plus the expected signal for a 95 GeV SM Higgs (dashed line). 
}
\vspace*{-1.cm}
\begin{center}
\begin{tabular}{l}\\
\epsfig{file=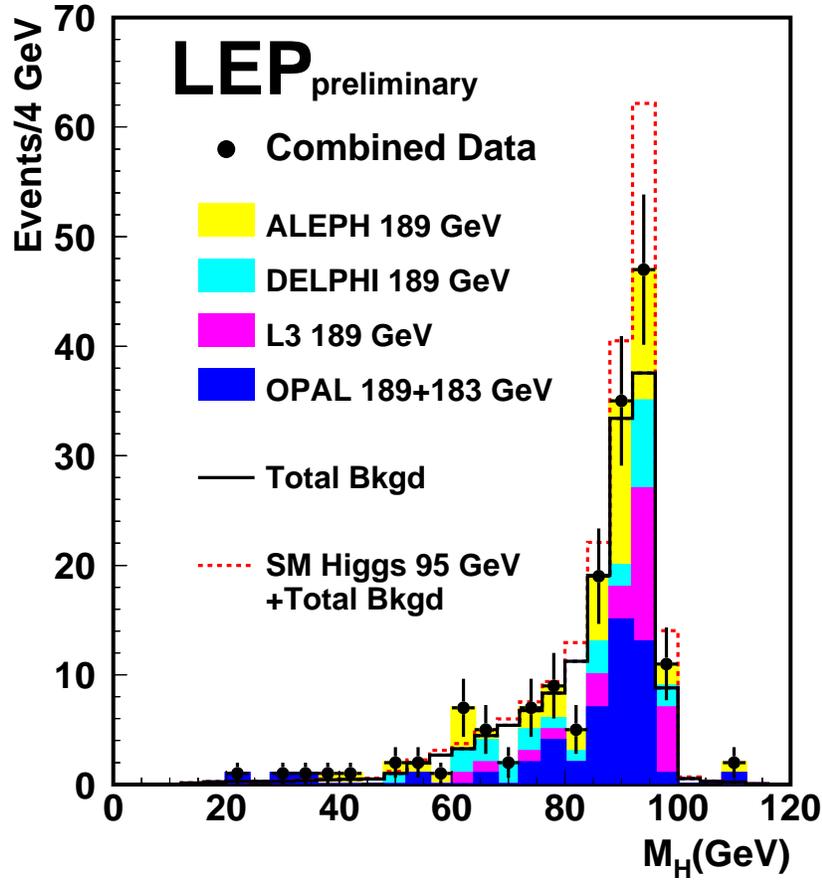,width=12.cm,clip=}\\
\end{tabular}
\end{center}
\end{figure}
\vspace*{-2.cm}

\begin{figure}[H]
\caption{Results of the SM Higgs search for OPAL \protect\cite{nuhgmo99}.
Left: limit on the production rate for the SM Higgs boson at 95\% CL
(solid line) compared to the number of expected signal events (dashed line) 
as a function of
the Higgs mass. Right: The observed confidence level for an hypothetical 
SM Higgs signal (solid line) and the expected average confidence 
level (dashed line) 
for the hypothesis that no signal, but only background, contributes 
to the observed candidates. 
} 
\vspace*{-2.cm}
\begin{center}
\begin{tabular}{ l l  }  \\ 
\hspace{-0.5cm}
\epsfig{file=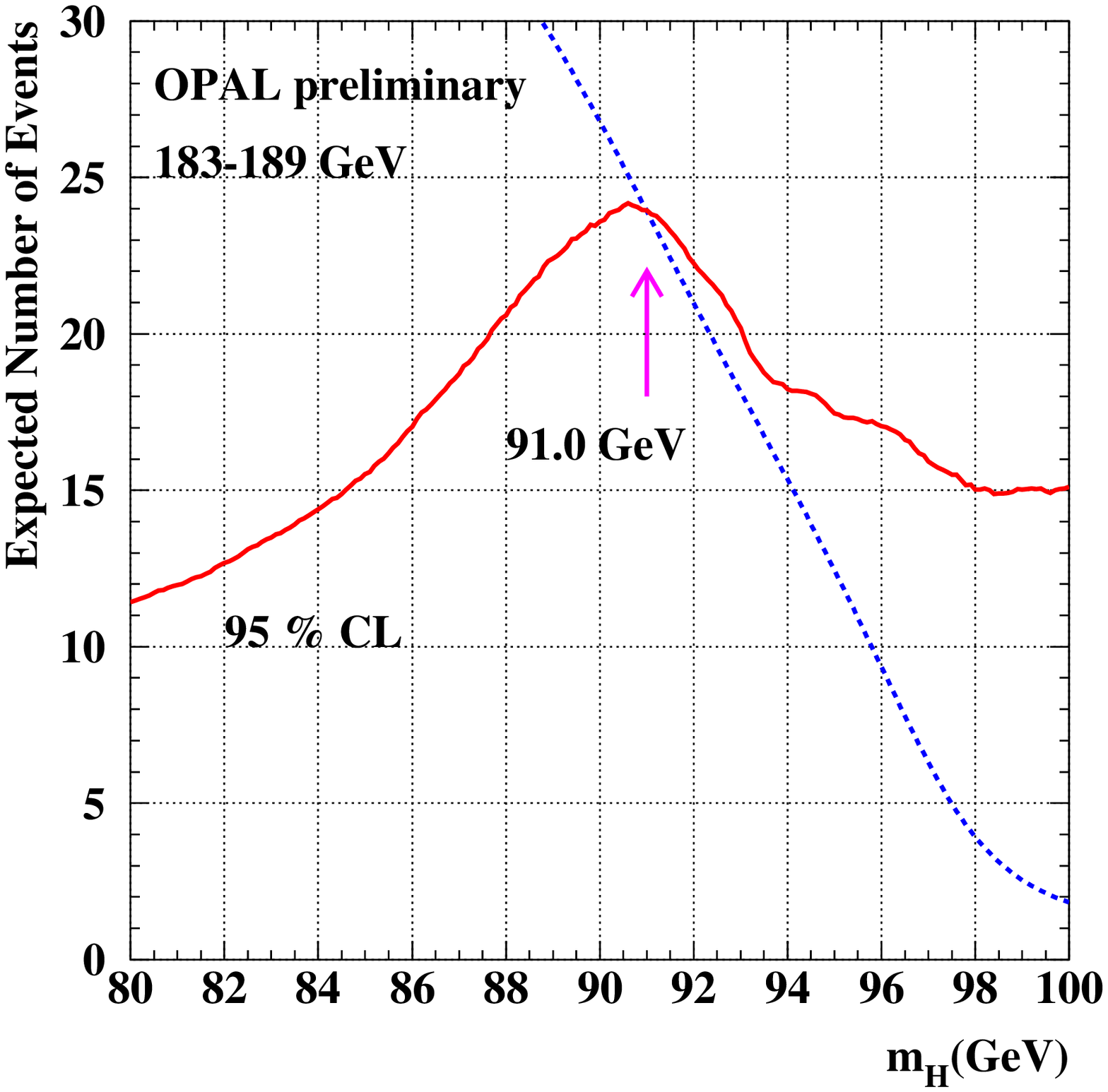,width=8.5cm,height=10.cm,clip=} &
\hspace{0.5cm}
\epsfig{file=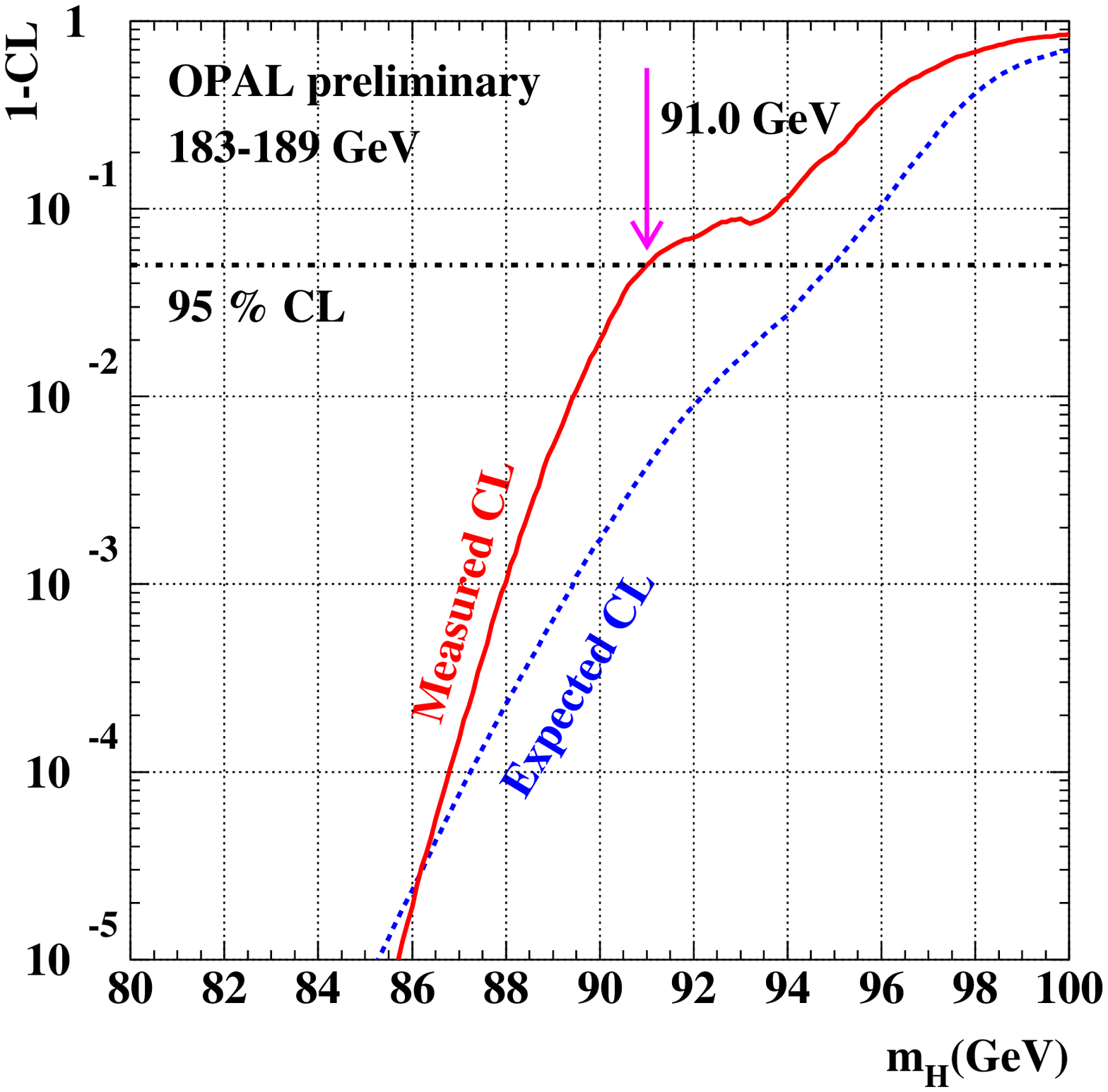,width=8.5cm,height=10.cm,clip=} \\
\end{tabular}
\end{center}
\end{figure}

\begin{table}[H]
\caption{Observed and expected lower limits on the SM Higgs mass together
with the probabilities to obtain a limit equal to or lower than the one 
actually observed \protect\cite{nuhgmo99}.
}
\bf
\begin{center}
\begin{tabular}{ l c c r  }  \\ 
\hline
              & Observed  &  Expected & Probability \\
\hline
 & & & \\
 ALEPH  &  90.2 GeV  &    95.7 GeV     & $\sim$ 1\%           \\
 DELPHI &  95.2 GeV  &    94.8  GeV    &  68\%             \\
 L3     &  95.2 GeV  &    94.4 GeV     &  55\%             \\
 OPAL   &  91.0 GeV  &    94.9 GeV     &   4\%           \\
\hline
\end{tabular}
\end{center}
\end{table}
\vspace*{-1.0cm}

\begin{figure}[H]
\caption{
Results of the MSSM Higgs searches. Left: Region (dark grey area) 
in the plane [$\Mh$,$\tanbet$] experimentally excluded 
by L3  for maximal mixing in the stop sector (resulting in more 
conservative limits than for no mixing); 
the light grey regions are not theoretically allowed,
in the hypothesis of maximal mixing; the dashed
(dotted-dashed) 
line indicates the expected average (median) lower limits on $\Mh$ 
as function of $\tanbet$ \protect\cite{nuhgmo99}. 
Right: Regions 
in the plane [$\MA$,$\tanbet$] experimentally excluded 
by DELPHI \protect\cite{nuhgmo99}, 
for maximal mixing (grey region) and for
no mixing (region on the left of the dotted line).  
}
\vspace*{-1.0cm}
\begin{center}
\begin{tabular}{ l l  }  \\ 
\hspace*{-1.5cm}
\epsfig{file=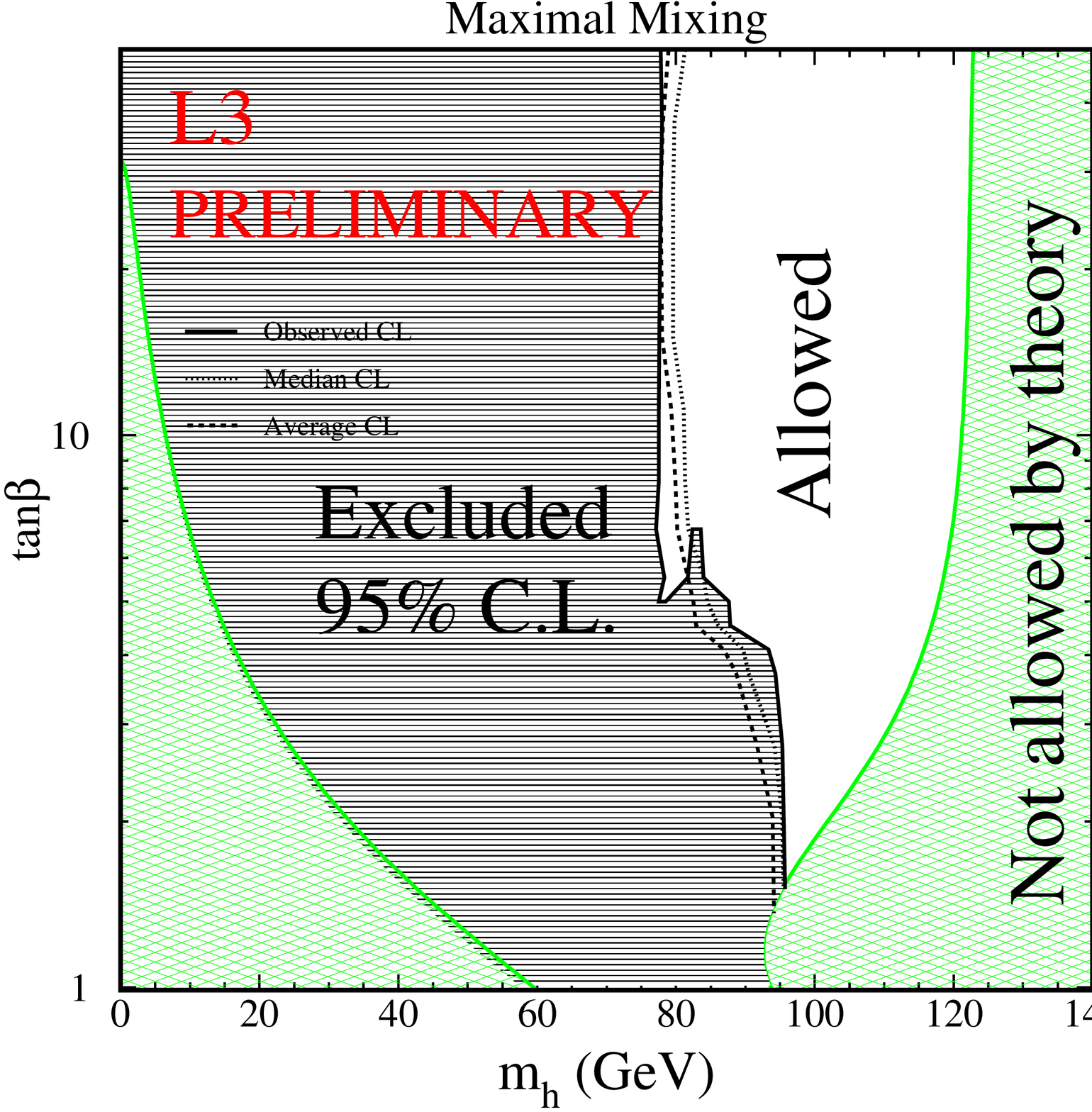,width=9.cm,height=10.cm,clip=} &
\hspace*{-0.5cm}
\epsfig{file=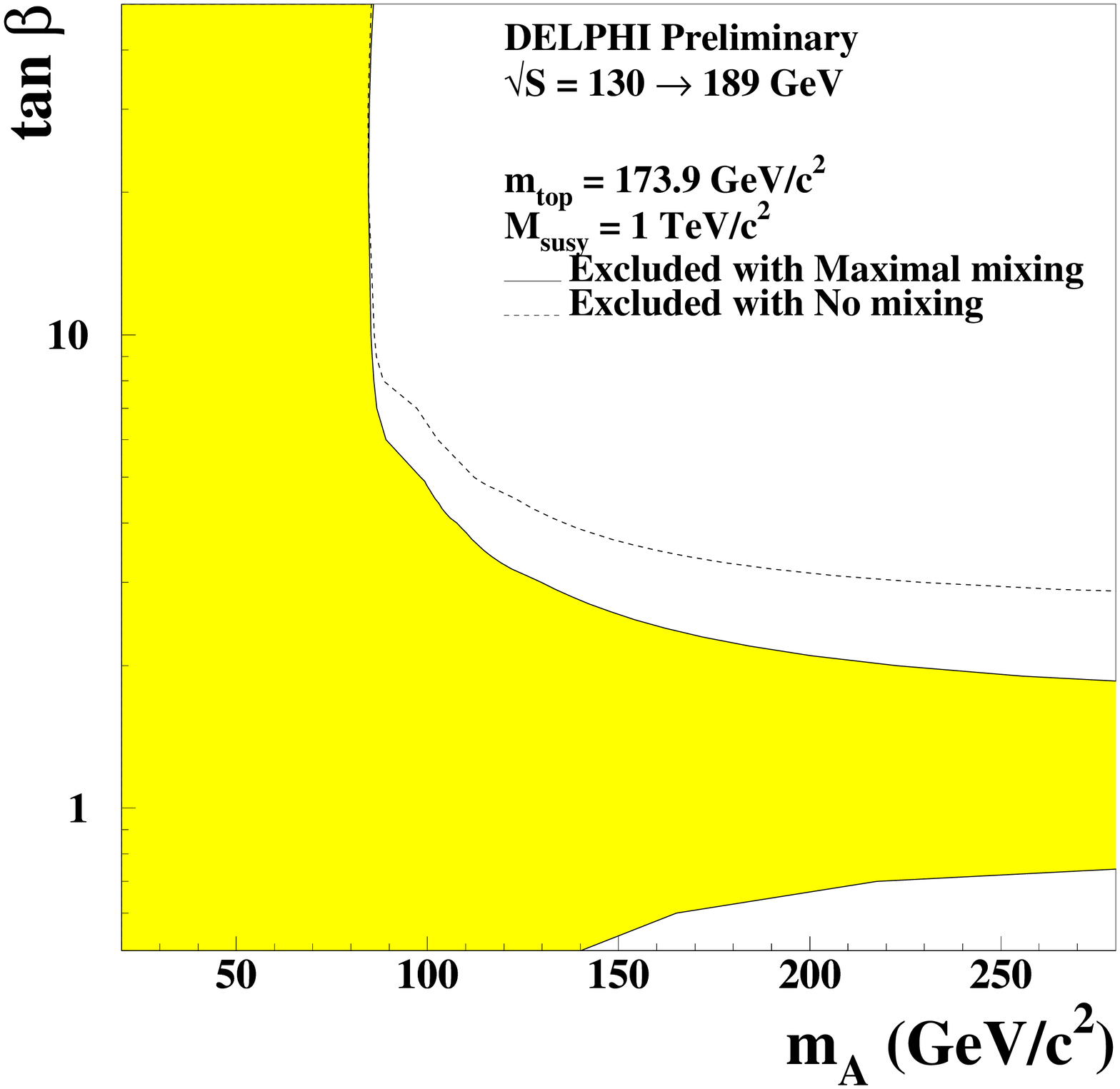,width=9.cm,height=10.5cm,clip=} \\
\end{tabular}
\end{center}
\end{figure}
\vspace*{-1.5cm}

\begin{table}[H]
\caption{Lower limits on the neutral MSSM Higgs masses 
and derived excluded ranges of $\tan{\beta}$ \protect\cite{nuhgmo99}, 
for the hypotheses 
of maximal mixing and no mixing in the stop sector. 
The numbers in parenthesis are the
expected average lower limits.}
\bf
\begin{center}
\begin{tabular}{ l c c c c  }  \\ 
\hline
              & \boldmath$m_h$(GeV)   & \boldmath$m_A$(GeV) & 
  \boldmath$\tan(\beta)$ 
& \boldmath$\tan{\beta}$ \\
 & & & max. mix. &  no mix. \\
\hline
ALEPH  &  80.8 (82.5)  & 81.2 (82.7)   
       & --            & 1$<\tan{\beta}<$2.2  \\
DELPHI & 83.5 (80.5)   & 84.5 (81.6) 
       & 0.9 $<\tan{\beta}<$ 1.5  & 0.6 $<\tan{\beta}<$2.6 \\
L3     & 77.0 (77.8) & 78.0 (77.9) 
& 1$<\tan{\beta}<$1.5  & 1$<\tan{\beta}<$2.6  \\
OPAL   & 74.8 (76.4) & 76.5 (78.2) 
       &  --         & 0.81$<\tan{\beta}<$2.19  \\
 & & & & \\
\hline
\end{tabular}
\end{center}
\end{table}
\begin{figure}[H]
\caption{
Charged Higgs mass lower limits \protect\cite{chhgmo99} 
as function of $BR(\Hpm\to\tau^\pm\nu)$ for ALEPH and OPAL. The experimentally
excluded regions  
in the plane $\MHpm$,$BR(\Hpm\to\tau^\pm\nu)$,
from the combination of the results in the different search channels, 
are shaded. The expected average lower limits are 
indicated by the dashed lines.}
\begin{center}
\begin{tabular}{ l l  }  \\ 
\hspace*{-1.cm}
\epsfig{file=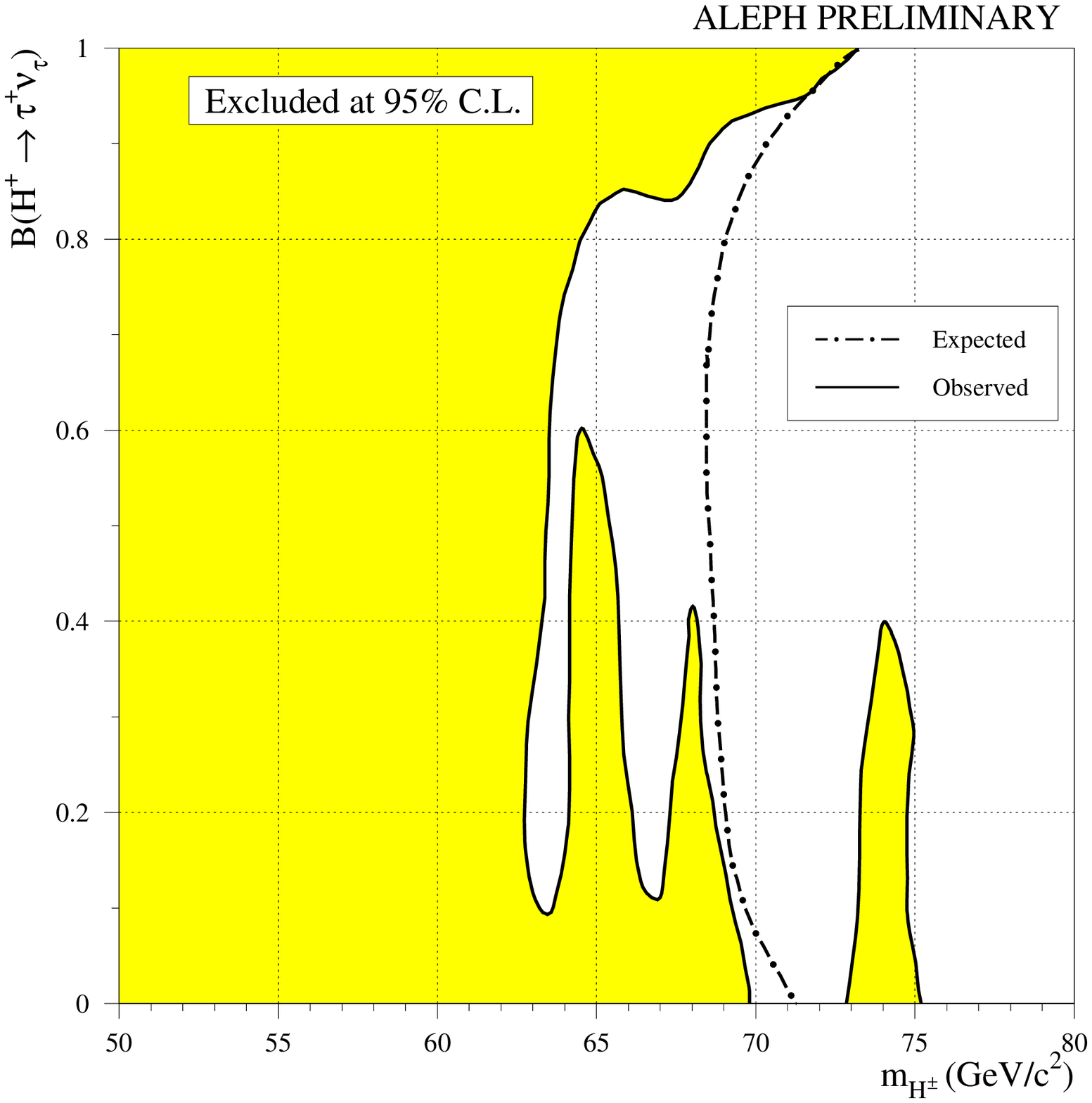,width=7.5cm,height=7.5cm,clip=} &
\hspace*{-0.1cm}\vspace*{1.cm}
\epsfig{file=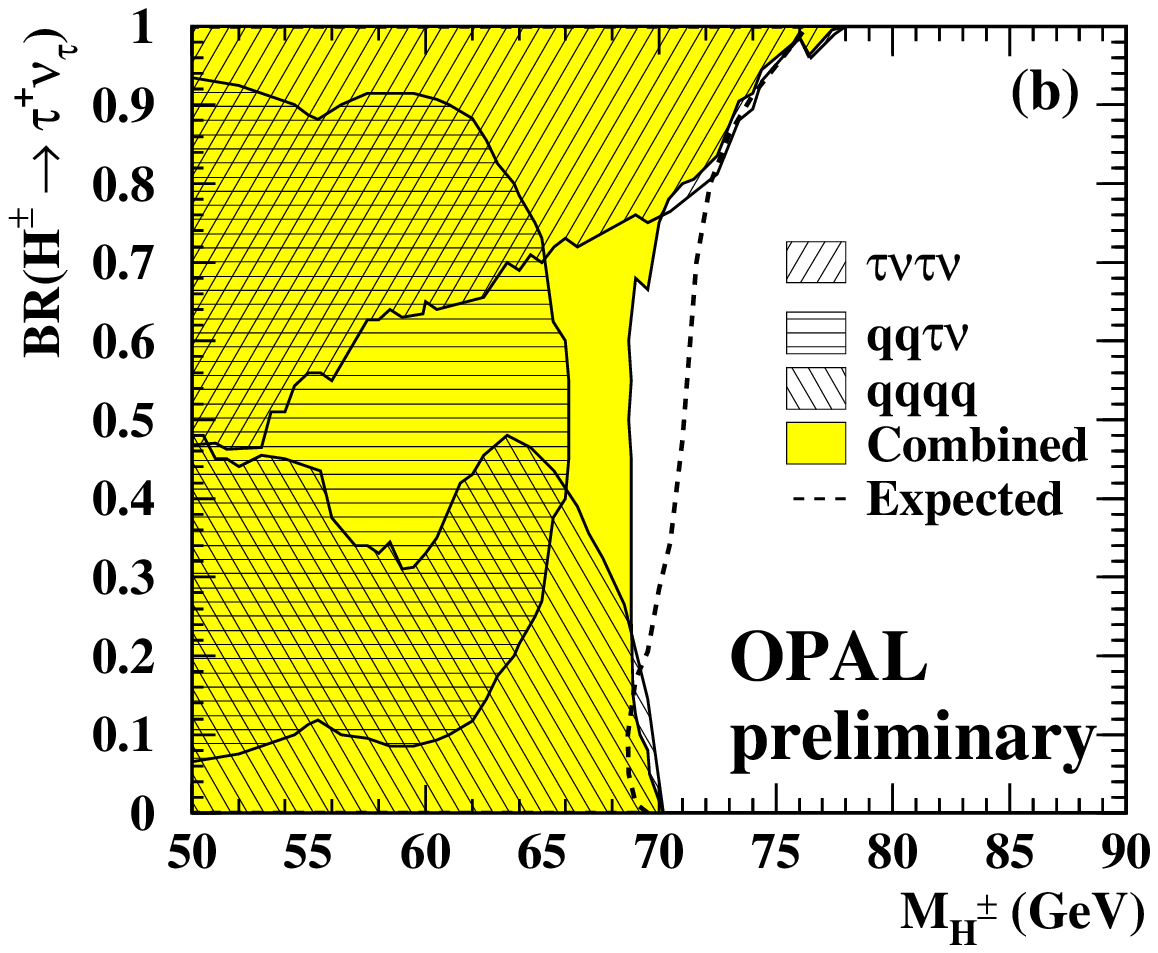,width=7.5cm,height=8.cm,clip=} \\
\end{tabular}
\end{center}
\end{figure}

\vspace*{-3.cm}

\begin{table}[H]
\begin{center}
\caption{Lower limits on the charged Higgs mass \protect\cite{chhgmo99}
for any value of $BR(\Hpm\to\tau^\pm\nu)$ (decay-mode independent), 
and for $BR(\Hpm\to\tau^\pm\nu)$=0 or 1. 
Number in parenthesis are the expected lower limits. 
}
\begin{tabular}{ l c c c  }  \\ 
\hline
BR$(H^\pm\rightarrow\tau\nu)=$ & any    & 0.   & 1. \\
\hline
                & $m_{H\pm}>$ & $m_{H\pm}>$ & $m_{H\pm}>$ \\ 
\hline
   ALEPH  & 62.5 (68.5)   & 69.8     &  73       \\
   DELPHI & 65.1 (69)     & 72       &  78       \\
    L3    & 67.5 (70)     & 67.5      & 78.5     \\
    OPAL  & 68.7 (67)     &  70      & 78        \\
\hline
\end{tabular}
\end{center}
\end{table}
\vspace*{-0.5cm}
\begin{figure}[H]
\caption{
Left: Higgs mass lower limits  
as function of  $BR\rm(\h\to inv))$ for DELPHI \protect\cite{inhgmo99}; 
the experimentally excluded region  
in the plane [$\Mh$,$BR\rm(\h\to inv)$],
from the combination of the results of the searches for visible (SM-like)
and invisible Higgs decays, 
is shaded. 
Right: experimentally excluded rates  
for $\HZ\to\gamma\gamma\gamma$,
$\rm\sigma(\Hv\gamma)\times BR(\Hv\to\gamma\gamma)$, compared to the  
expected rates in the presence of anomalous Higgs
couplings at $\sqrts=$ 189 GeV \protect\cite{gghgmo99}. 
The expected rates are plotted for 
different values of $\Lambda$, the typical energy scale at which the 
new interactions should take place. 
}
\vspace*{-0.5cm}
\begin{center}
\begin{tabular}{ c c  }  \\ 
\hspace*{-1.cm}
\epsfig{file=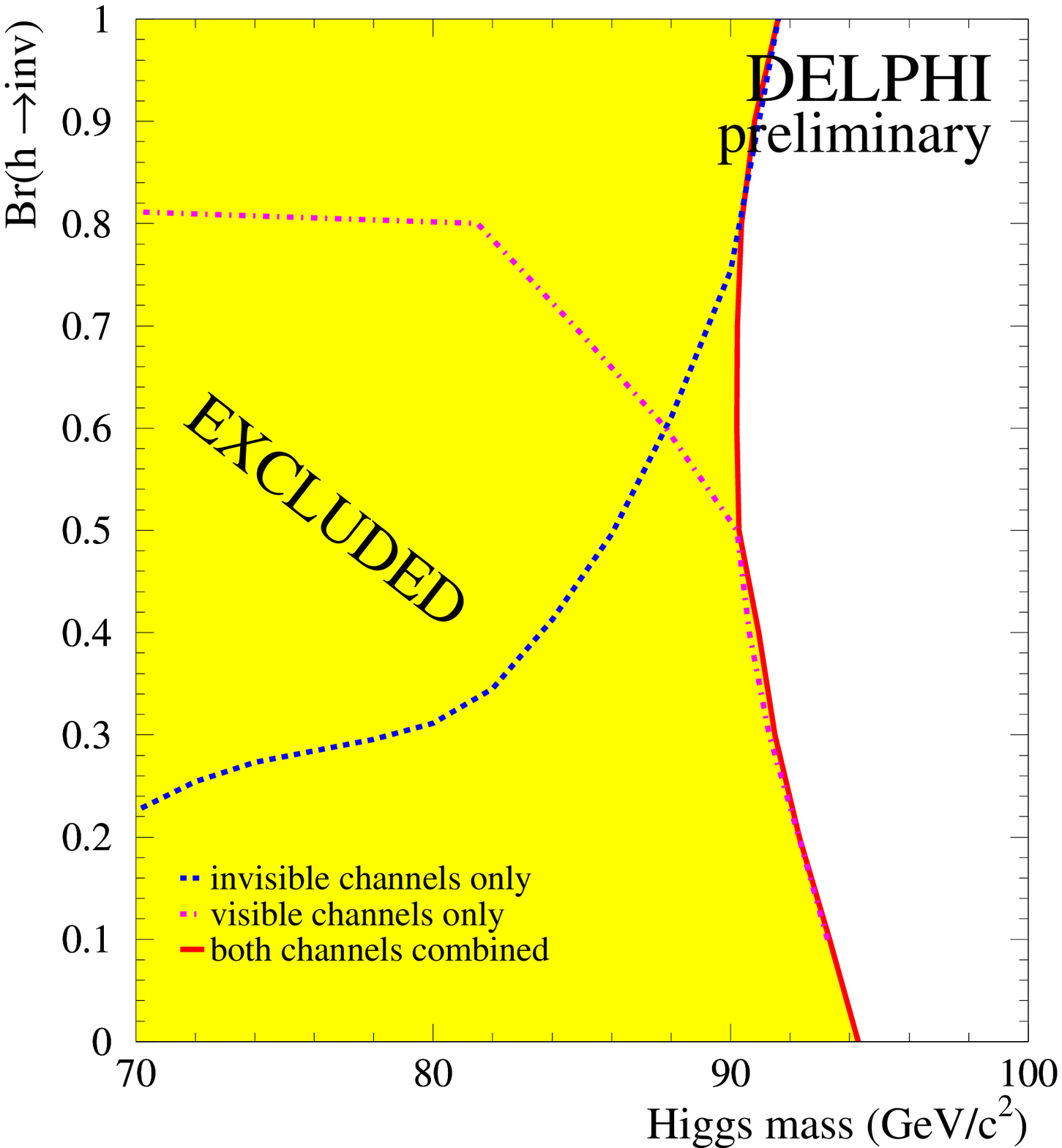,width=8.cm,height=7.cm,clip=} &
\hspace*{0.5cm}
\epsfig{file=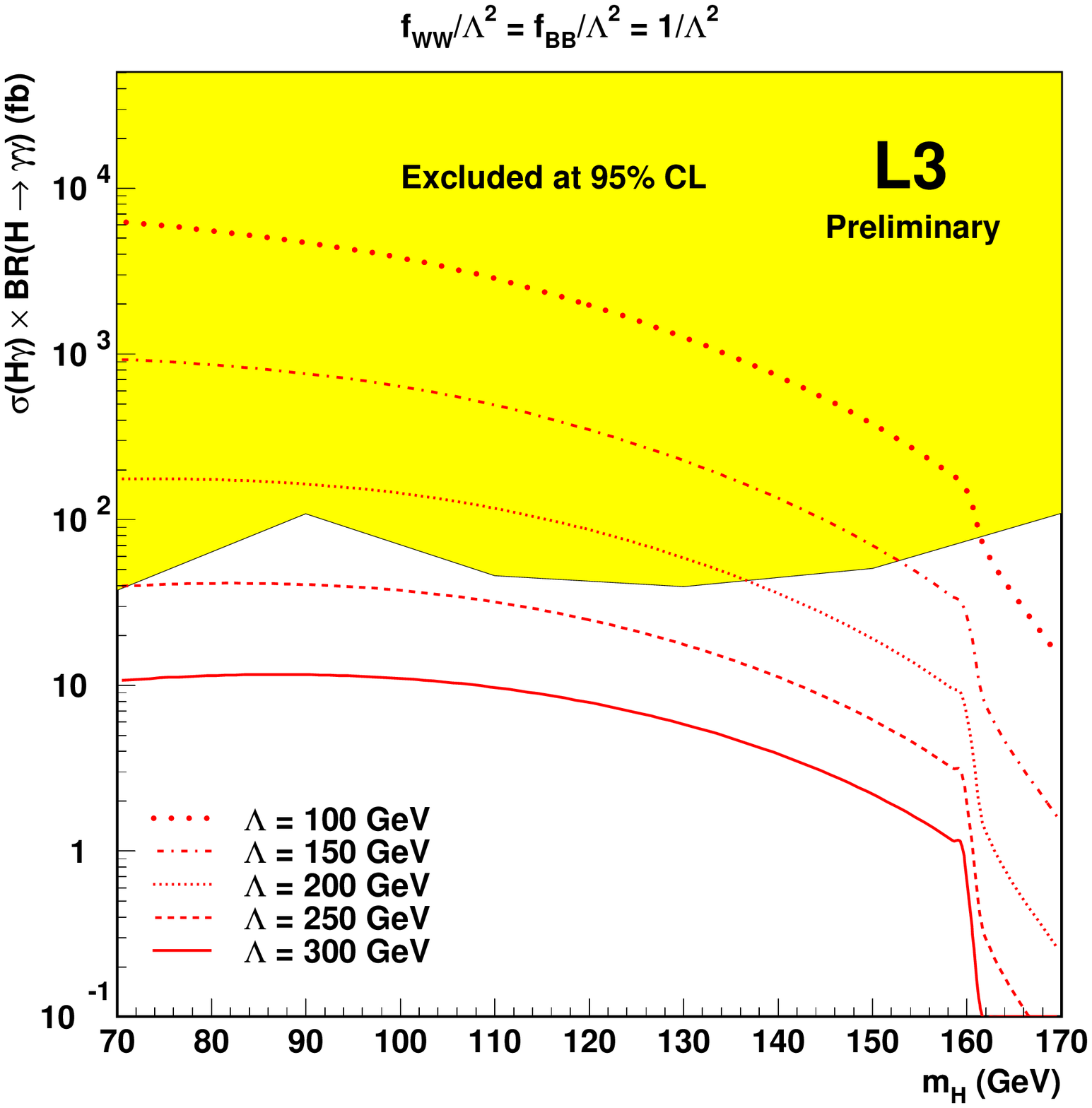,width=8.cm,height=7.5cm,clip=} \\
\\
\end{tabular}
\end{center}
\end{figure}

%% file: papeth.bbl
\section*{References}
\begin{thebibliography}{99}

\bibitem{lep2yebo} Report CERN 96-01, Vol. 1, ``Physics at LEP2'', edited by 
G. Altarelli, T. Sj\"{o}strand and F. Zwirner and references therein.

\bibitem{heinson99} A. Heinson, these proceedings. 

\bibitem{lemo99ew}
LEP Electroweak Working Group, Results for the Winter Conferences in 
{\tt http://www.cern.ch/LEPEWWG/plots/winter99/}

\bibitem{lemo99pp}
LEP Working Group for Higgs boson searches, CERN-EP/99-060 and references 
therein.  

\bibitem{cavallari99} F. Cavallari, these proceedings and references therein. 

\bibitem{nuhgmo99}
ALEPH Coll., ALEPH 99-007 CONF 99-003,
DELPHI Coll., DELPHI 99-8 CONF 208,
L3 Coll., L3 note 2382 and  L3 note 2383, 
OPAL Coll., OPAL PN382, March 1999.

\bibitem{mass} The unpublished 189 GeV data for the mass spectrum combination 
were kindly provided by the four LEP Collaborations. In particular, I wish to 
thank P. McNamara and P. Gay (ALEPH), W. Murray and P. Lutz (DELPHI) and 
K. Desch (OPAL) for providing me with informations and explanations. 


\bibitem{chhgmo99}
ALEPH Coll., ALEPH 99-016 CONF 99-011, 
DELPHI Coll., DELPHI 99-17 CONF 217,  
L3 Coll., L3 note 2379, 
OPAL Collaboration, OPAL PN373, March 1999. 
 
\bibitem{inhgmo99}
ALEPH Coll., ALEPH 99-013 CONF 99-008, 
DELPHI Coll., DELPHI 99-16 CONF 216, March 1999, and references therein. 

\bibitem{opgg98pa}
   OPAL Collaboration,  
   \PLB\ 437 (1998) 218 and references therein. 

\bibitem{gghgmo99}
L3 Coll., L3 note 2365, March 1999.

\bibitem{degg99ep}
DELPHI Collaboration, 99--58 (April 1999), 
   accepted by \PLB , and references therein . 

\end{thebibliography}
